\documentclass[aps,prb,twocolumn,superscriptaddress,showpacs]{revtex4}
\pdfoutput=1

\usepackage{graphicx}
\usepackage{amsmath}
\usepackage{color}

\newcommand{\dg}{\dagger}
\newcommand{\la}{\langle}
\newcommand{\ra}{\rangle}

\begin{document}

\title{Influence of Coulomb correlations on the quantum well intersubband absorption at low temperatures}

\author{Thi Uyen-Khanh Dang}
\email[]{uyen@itp.physik.tu-berlin.de}
\author{Carsten Weber}
\affiliation{Institut f\"ur Theoretische Physik, Nichtlineare Optik und Quantenelektronik, Technische Universit\"at Berlin, Hardenbergstr. 36, 10623 Berlin, Germany}
\author{Marten Richter}
\affiliation{Institut f\"ur Theoretische Physik, Nichtlineare Optik und Quantenelektronik, Technische Universit\"at Berlin, Hardenbergstr. 36, 10623 Berlin, Germany}
\affiliation{Department of Chemistry, University of California Irvine, Irvine, California 92697, USA}
\author{Andreas Knorr}
\affiliation{Institut f\"ur Theoretische Physik, Nichtlineare Optik und Quantenelektronik, Technische Universit\"at Berlin, Hardenbergstr. 36, 10623 Berlin, Germany}

\date{\today}

\begin{abstract}
We present a theory for the intersubband absorption including electronic ground-state correlations in a doped GaAs/Al$_{0.35}$Ga$_{0.65}$As quantum well system.
Focusing on the influence of the Coulomb
interaction among the carriers at low temperatures, we find that the ground-state correlations lead to an increased renormalization and spectral broadening of the absorption spectrum. At $T$ = 1 K, its full width at half maximum is increased by up to a factor 3. The inclusion of electron-phonon scattering strongly reduces the relative impact of the electronic correlations.
\end{abstract}


\maketitle

As a source of Terahertz and infrared radiation, semiconductor quantum well intersubband (ISB) transitions have been widely investigated experimentally.
\cite{Elsaesser:PhysRep:99,Williams:NaturePhotonics:07} 
In particular, an understanding of the spectral linewidth is of crucial importance for the design and control of heterostructure\cite{Helm:PhysRevLett:89} and laser devices.\cite{Chow:ApplPhysLett:97} 
Although the theoretical description of ISB dynamics has been the subject of many studies,
\cite{Iotti:PhysRevLett:01,Butscher:PhysStatusSolidiB:04,Pereira:PhysRevB:04,Li:PhysRevB:04,Waldmuller:PhysRevB:04,Zheng:ApplPhysLett:04,Shih:PhysRevB:05,Kira:PhysRevA:06,Waldmueller:IEEEJQuantumElectron:06,Butscher:PhysRevLett:06,Pasenow:SolStateComm:08,Weber:PhysRevB:09,Vogel:PhysRevB:09}
 current models for the ISB optics become insufficient at low temperatures, yielding an intrinsic absorption linewidth which is much smaller than experimentally observed.\cite{Kaindl::00,Waldmuller:PhysRevB:04,Shih:PhysRevB:05} For an equilibrium Fermi distribution of the electrons in the ground-state, Pauli blocking prevents electron-electron scattering at low temperatures, while the electron-phonon contribution is too weak to explain the experimental findings. To approach the experimental findings in the low temperature regime, some theoretical descriptions focus on disorder corrections such as impurities\cite{Banit:ApplPhysLett:05} or interface roughness.\cite{Li:PhysRevB:04}

In this article, we demonstrate that even in the absence of disorder, correlations in the electronic ground-state of the doped quantum well yield a line broadening at low temperatures not observed before:
Coulomb correlations lead to modified distribution functions which partially reduce the strong Pauli blocking in the low temperature regime by opening up new electron-electron scattering channels.\cite{Takada:PhysRevB:91,Daniel:PhysRev:60} This leads to additional dephasing which manifests itself in an enhanced broadening of the absorption spectrum.
Even if the additional inclusion of electron-phonon scattering strongly reduces the influence of the correlations, a thorough investigation of ground-state correlations clearly improves our physical understanding of the many-body effects in high quality quantum well samples.

Our approach is as follows: (I) after summarizing the overall dynamics of a two-dimensional electron gas in an ISB system, (II) we discuss the origin and (III) aspects of ground-state correlations and finally show their impact on the ISB absorption spectrum.

\begin{figure}[t]
\centering
\includegraphics[clip, width=0.5\columnwidth]{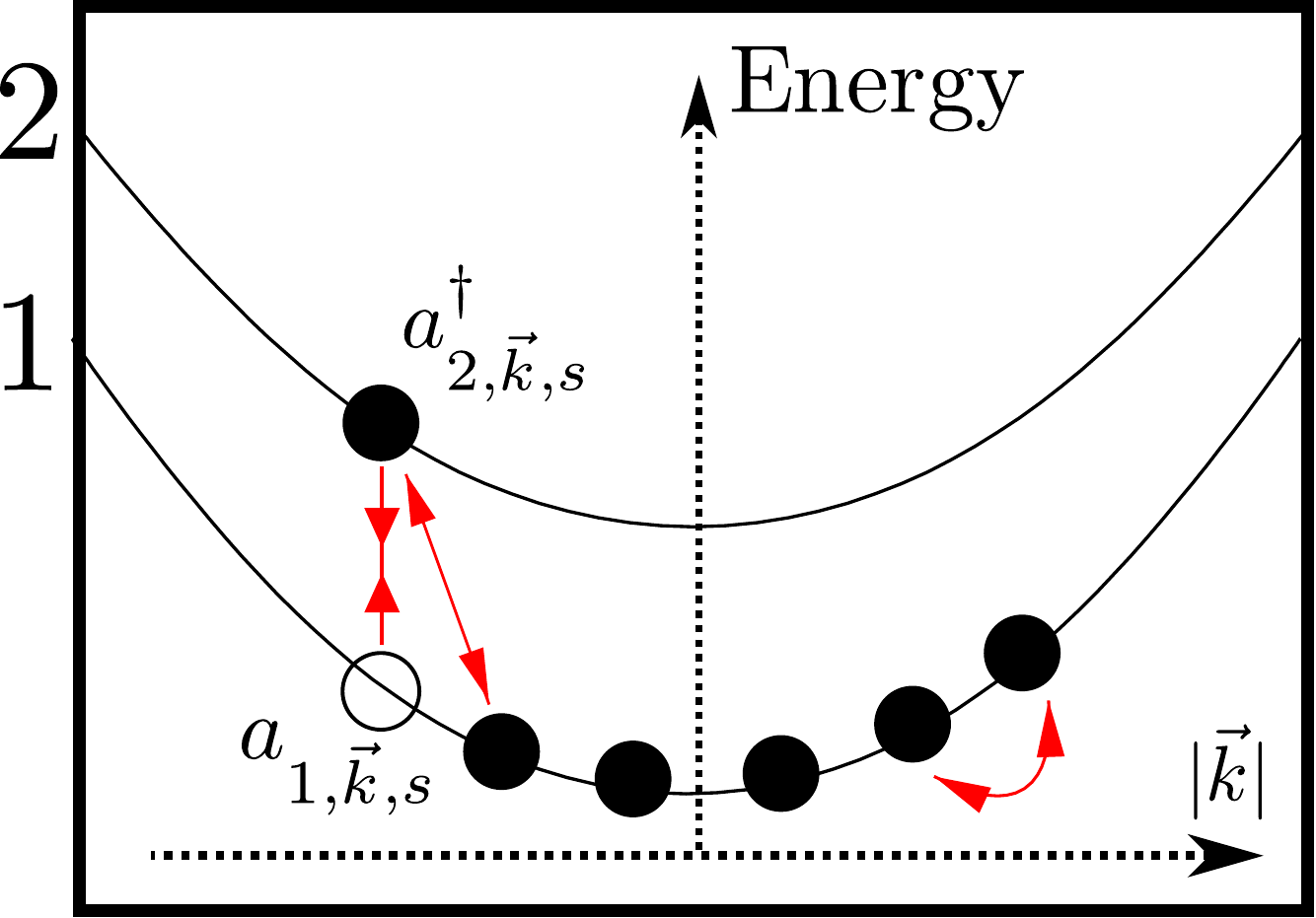}
\caption{(Color online) Sketch of the two energetically lowest conduction subbands of an n-doped  quantum well. The arrows symbolize the different Coulomb many-body interactions among the electrons. Optical excitation creates an electron ($a^{\dagger}_{2, \vec k, s}$) in band 2 and annihilates an electron ($a^{{}}_{1, \vec k, s}$) in band 1.}
\label{fig:sketch}
\end{figure}

(I) {\it ISB dynamics.} For our investigations, we consider an n-doped $\text{GaAs}/\text{Al}_{0.35}\text{Ga}_{0.65}\text{As}$ quantum well system where only the kinetics of the two lowest conduction subbands are of relevance.
\footnote{For the calculations, we used the following parameters: static dielectric constant $\varepsilon_0$ = 12.9, high-frequency dielectric constant $\varepsilon_\text{bg}$ = 10.9, longitudinal optical phonon energy $\hbar \omega_{\rm LO}$ = 36 meV, subband gap $\varepsilon_{\rm gap}$ = 210.66 meV, effective masses $m_1^* = 0.078 m_0$, $m_2^* = 0.131 m_0$.} 
This is a good approximation since we are focusing on low temperatures and subband gaps which are typically quite large ($k_\text{B}T \lesssim $ 9 meV $\ll$ 100 meV $\lesssim \varepsilon_\text{gap}$ ). To approximate the more realistic finite potential of the quantum well, we use an effective well width for an infinite potential well.\cite{Waldmuller:PhysRevB:04} Non-parabolicity effects are included in the form of different effective subband masses.\cite{Ekenberg:PhysRevB:89} A sketch of the considered quantum well in-plane band structure is given in Fig.~\ref{fig:sketch}.

The Hamiltonian of the investigated system consists of the in-plane kinetics of the confined electrons ($H_{0,\text{el}}$), the Coulomb interaction ($H_\text{C}$), and the semiclassical coupling to an external light field ($H_\text{em}$): 
\begin{align}
H_{0,\text{el}} &= \sum_{i,\vec k_i, s_i} \varepsilon_{i,\vec k_i}a^{\dagger}_{i, {\vec k_i}, s_i} a^{\phantom{\dagger}}_{i, {\vec k_i}, s_i} ,\label{eq:H_0,el} \\
H_\text{C} &= \frac{1}{2} \sum_{\{i j l m\}} V^{|\vec k{i} - \vec k_{l}|}_{\{{i j l m}\}} \, a^{\dagger}_{\{i\}} a^{\dagger}_{\{j\}} a^{}_{\{m\} } a^{}_{\{l\}} ,  \label{eq:H_C}\\
H_\text{em} &= \sum_{i, j, \vec k_i, s_i} \hbar \Omega_{\rm em}(t)a^{\dagger}_{i, {\vec k_i}, s_i} a^{\phantom{\dagger}}_{j, {\vec k_i}, s_i}\label{eq:H_em} .
\end{align} 
Here, $a^{\dagger}_{i, \vec k_i, s_i}$ ($a^{\phantom{\dagger}}_{i, \vec k_i, s_i}$) denotes the creation (annihilation) operator for an electron in subband $i$ with an in-plane wave vector ${\vec k_i}$, spin $s_i$, and energy $\varepsilon_{i, \vec k_i} = \varepsilon_i + (\hbar^2 {\vec k^2_i}/2m_i)$ (cf. Fig.~\ref{fig:sketch}).  We introduce the compound index $\{i\} = \{i, \vec k_i, s_i\}$ to simplify the notation.
Furthermore, $V^{|\vec k{i} -\vec k_{l}|}_{\{{i j l m}\}} = V^{|\vec k{i} -\vec k_{l}|}_{{i j l m}} \delta_{\vec k_i + \vec k_j,\vec k_l + \vec k_m }\delta_{s_i,s_l} \delta_{s_j,s_m}$ describes the Coulomb-induced transitions of electrons in the states $\{l,m\}$ to the states $\{i,j\}$, where $V^{|\vec k{i} -\vec k_{l}|}_{{i j l m}}$ is the Coulomb matrix element.\cite{Waldmuller:PhysRevB:04}
The Rabi frequency $\Omega_{\rm em}(t) = {\bf d}_{12} \cdot {\bf E}(t)/\hbar$ (dipole moment ${\bf d}_{12}$) describes the interaction between the external field ${\bf E}(t)$ and the electronic system.  
The absorption coefficient $\alpha(\omega)$ of the ISB system is calculated via the complex susceptibility $\chi(\omega)$ as $\alpha(\omega) \propto \omega\text{Im}\left[\chi(\omega)\right]$, $\chi(\omega) = P(\omega)/\epsilon_0 E(\omega)$ with the macroscopic polarization 
$P(\omega)$ determined by the microscopic ISB polarizations $\rho_{ij, \vec k, s} = \langle a^{\dagger}_{i, {\vec k}, s} a^{\phantom{\dagger}}_{j, {\vec k}, s}\rangle$ $ (i\neq j)$ [cf. Eq.~\eqref{eq:polariz}].\cite{Schafer::02}

The calculations of the ISB dynamics for the polarizations $\rho_{i j,\vec k, s}$ are carried out within a density-matrix approach using a correlation expansion.\cite{Lindberg:PhysRevB:88,Rossi:RevModPhys:02} Evaluating the system up to second order, the dynamical equations for the polarizations 
and the electronic populations $f_{i, {\vec k}, s} = \langle a^{\dagger}_{i, {\vec k}, s} a^{\phantom{\dagger}}_{i, {\vec k}, s}\rangle$ as well as the second-order correlations  
$\sigma^\text{c}_{\{i j l m\}} = \langle a^{\dagger}_{\{i\}} a^{\dagger}_{\{j\}} a^{\phantom{\dagger}}_{\{l\}} a^{\phantom{\dagger}}_{\{m\}}\rangle - (\langle a^{\dagger}_{\{i\}}a^{\phantom{\dagger}}_{\{m\}} \rangle \langle a^{\dagger}_{\{j\}}a^{\phantom{\dagger}}_{\{l\}} \rangle - \langle a^{\dagger}_{\{i\}}a^{\phantom{\dagger}}_{\{l\}} \rangle \langle a^{\dagger}_{\{j\}}a^{\phantom{\dagger}}_{\{m\}} \rangle )$ 
are coupled by the electron-electron interaction. 
In the following, we will neglect the spin index since the quantities of interest are spin independent for our system.

Evaluating the dynamical equations for the electronic polarizations between the two lowest subbands $\rho_{1 2, \vec k}$ with the Hamiltonian given above yields
\begin{align}
\frac{\text{d}}{\text{dt}} \rho_{ 1 2, {\vec k}} &= \frac{i}{\hbar} \la[H_{0,\text{el}} + H_\text{em} + H_\text{C}, a^{\dg}_{1, {\vec k}} a^{\phantom{\dg}}_{2, {\vec k}}]\ra \notag\\
 &= 
- \frac{i}{\hbar} (\tilde{\mathcal{E}}_{2,\vec k} -\tilde{\mathcal{E}}_{1,\vec k} )\rho_{12, \vec k}- i \tilde{\Omega}(t) (f_{2,\vec k} -f_{1,\vec k}) 
\notag \\
&\quad - \Gamma\{\rho_{i j,\vec k},f_{i, \vec k}\} . \label{eq:polariz}
\end{align}
The energy $\tilde{\mathcal{E}}_{i,\vec k}$ 
combines the subband energy $\varepsilon_{i,\vec k}$ and the energetic renormalization due to the Coulomb exchange contribution.
The renormalized Rabi frequency $\tilde{\Omega}(t)$ 
contains the Rabi frequency of the external light field $\Omega_{\rm em}(t)$ and a renormalization due to the Coulomb exciton and depolarization contributions.\cite{Li:PhysRevB:04,Waldmuller:PhysRevB:04} 
These renormalizations result from a Hartree-Fock approximation, which is first order in the Coulomb potential.\cite{Chuang:PhysRevB:92,Nikonov:PhysRevB:99}
The second-order correlation contribution yields the dephasing functional $\Gamma\{\rho_{i j,\vec k},f_{i, \vec k}\}$ caused by Boltzmann scattering between the electrons and includes both diagonal and nondiagonal Coulomb scattering contributions which lead to a broadening of the absorption spectrum. 
\footnote{While we include all diagonal terms (proportional to $\rho_{ij,\vec k}$) in the kinetic scattering contributions, we only consider the nondiagonal terms in Ref.~\onlinecite{Waldmuller:PhysRevB:04} which are proportional to $\rho_{ij,\vec k - \vec q}$ since these are the dominant terms counteracting the broadening of the diagonal terms.} Depending on the nonparabolicity of the bandstructure, a strong cancellation between diagonal and nondiagonal terms can occur.\cite{Li:PhysRevB:04} For further details concerning the Hartree-Fock and kinetic scattering contributions, see Ref.~\onlinecite{Waldmuller:PhysRevB:04}.

Additional higher-order correlations are included via a phenomenological damping $\gamma$ of the second-order correlation functions,\cite{Schilp:PhysRevB:94} which softens the strict energy conservation in $\Gamma\{\rho_{i j,\vec k},f_{i, \vec k}\}$ typically used in Markovian approaches. 
In particular, $\gamma$ enters $\Gamma\{\rho_{i j,\vec k},f_{i, \vec k}\}$ and the correlation correction of the ground-state distribution $\delta f_{ {\vec k}}$ (discussed in the next section) in a way that a strong cancellation of the influence of $\gamma$ on the spectral broadening occurs.\footnote{The deviation $\delta f_{ {\vec k}}$  increases for decreasing $\gamma$ while the linewidth of the absorption spectrum neglecting correlation contributions, determined by $\Gamma\{\rho_{i j,\vec k},f_{i, \vec k}\}$, is reduced for decreasing $\gamma$. Thus, the broadening effect of $\delta f_{ {\vec k}}$ on the absorption spectrum is always in a reasonable proportion to the reduction of the linewidth so that the influence of $\gamma$ on the spectral broadening is strongly canceled.}
For the purpose of this paper, it is thus justified to focus on a fixed value of $\hbar\gamma$ which is chosen as 5~meV here.

Due to Pauli-blocking in the dephasing functionals $\Gamma\{\rho_{i j,\vec k},f_{i, \vec k}\}$, the broadening of the spectrum becomes very narrow for low temperatures due to the sharp Fermi edge. This result, which is not in agreement with experimental findings, leads us to the assumption that some fundamental suppositions in the theory must be corrected. Here, we address additional ground-state correlations neglected so far:

(II) {\it Ground state correlations.}
In most theoretical descriptions of $\tilde{\mathcal E}_{i, \vec{k}}$, $\tilde{\Omega}$, and $\Gamma$, the electronic populations $f_{i, \vec k}$ in equilibrium are taken to be Fermi distributions $f^{(0)}_{i, {\vec k}}$ of the non-interacting electron gas.\cite{Butscher:PhysStatusSolidiB:04,Li:PhysRevB:04,Waldmuller:PhysRevB:06} This yields good agreement with the experiment in the high temperature regime.\cite{Kaindl:PhysRevLett:98,Li:PhysRevB:04,Waldmuller:PhysRevB:04} At low temperatures, the electron-electron ground-state correlations are expected to play an important role for the following reason: While for high temperatures the kinetic energy of the electrons clearly dominates over the Coulomb repulsion, the latter becomes more important for low temperatures where the kinetic energy and the Coulomb repulsion may be of a similar magnitude.

To extract the correlation of the ground-state beyond the usual second-order Born-Markov approximation, we  consider a deviation $\delta f_{i, {\vec k}} = f_{i, {\vec k}} - f^{(0)}_{i, {\vec k}}$ from the equilibrium Fermi distribution $f^{(0)}_{i, {\vec k}}$ (cf. Refs.~\onlinecite{Takada:PhysRevB:91,Daniel:PhysRev:60} for a treatment of ground-state correlations in metals) and include first-order memory effects. The evolution of $\delta f_{i, {\vec k}}$ is described by the kinetics $H_\text{0,el}$ of the non-interacting electron gas as well as the Coulomb coupling $H_{\text{C}}$:
\begin{align}
- i \hbar \frac{\text{d}}{\text{dt}} \delta f_{i, {\vec k}} = - i \hbar \frac{\text{d}}{\text{dt}} f_{i, {\vec k}} = \la[H_{0,\text{el}}+H_\text{C},a^{\dg}_{i,{\vec k}} a^{{\phantom{\dg}}}_{i, {\vec k}}]\ra. \label{eq:H_gs}
\end{align}
We evolve Eq.~\eqref{eq:H_gs} up to second order in the Coulomb coupling similar to the derivation of the ISB polarization. Restricting the correlation effects to a single subband (ground state), we neglect the subband index $i$ in the following. In this case, the first-order correlation contributions vanish, and in second order, we obtain linear differential equations for $\sigma^\text{c}_{\{i j l m\}}$ with a time dependent inhomogeneity $Q(t)$:
\begin{align}
- i \hbar \frac{\text{d}}{\text{dt}} \delta f_{{\vec k}} 
&= \sum_{\vec k', \vec q} 
\left[\sigma^\text{c}_1({\vec k},{\vec q},{\vec k'}) - \sigma^\text{c}_2({\vec k},{\vec q},{\vec k'})\right] 
\tilde{V}^{|{\vec q}|} ,\label{eq: dev_diff}\\
- i \hbar \frac{\text{d}}{\text{dt}} \sigma^\text{c}_{1/2} 
&= (\pm \Delta \varepsilon + i\hbar \gamma) \sigma^\text{c}_{1/2} \pm \tilde{W}^{|{\vec k,\vec q,\vec k'}|}Q(t), \label{eq:sigma_c_diff}
\end{align}
with the abbreviations $\tilde{W}^{|{\vec k,\vec q,\vec k'}|}=2\tilde{V}^{|{\vec q}|} - \tilde{V}^{|{\vec k' - \vec q - \vec k}|}$ and $\Delta \varepsilon= \varepsilon_{\vec{k} - \vec{q}} +\varepsilon_{\vec{k'}+\vec{q} } -\varepsilon_{\vec{k'} } -\varepsilon_{\vec{k} }$,
where $\tilde V$ denotes the screened Coulomb matrix element of the lower subband due to the modification of the potential by the presence of the other electrons.\cite{Lee:PhysRevB:99} Integrating Eq.~(\ref{eq:sigma_c_diff}), and assuming the memory of $Q(t)$ to be small, we can expand the inhomogeneity $Q(t')$ in a perturbation series around the local time $t$, yielding the following expression for $\sigma^\text{c}_{1/2}$:
\begin{align}
 \sigma^\text{c}_{1/2} = \mp \frac{i}{\hbar} \int_0^{\infty} \mathrm{d}s \, e^{(\pm \frac{i}{\hbar}\Delta \varepsilon - \gamma) s} \tilde{W}^{|{\vec k,\vec q,\vec k'}|}\left[ Q(t) + s \ \dot{Q}(t)\right]. \label{eq:sigma_c_int}         
\end{align}
The zeroth-order term $\propto Q(t)$ yields the typical Boltzmann scattering contributions which vanish for the Fermi distribution functions.
\footnote{While this term vanishes exactly for $\gamma = 0$, it yields a small but finite value for finite values of $\gamma$ which is neglected here.} The second term includes memory effects in first order (containing the temporal derivative of the source) and leads to
\begin{align}
 \sigma^\text{c}_{1/2} =   \mp \frac{i}{\hbar}\tilde{W}^{|{\vec k,\vec q,\vec k'}|}\dot{Q}(t)\left[ \frac{1}{(\pm \frac{i}{\hbar} \Delta \varepsilon - \gamma)^2} \right] ,
\end{align}
which is substituted in Eq.~\eqref{eq: dev_diff}.

In a first order iteration, we take the electron occupations occuring in the inhomogeneity $Q(t)$ to be Fermi distributions. 
This leads to the final expression of the deviation $\delta f_{{\vec k}}$:
%
\begin{align}
\delta f_{\vec k} &= 2 \sum_{\vec k', \vec q}\frac{\Delta \varepsilon^2 - \hbar^2\gamma^2}{(\Delta \varepsilon^2 + \hbar^2\gamma^2)^2} \left( \tilde V^{|{\vec k}'-{\vec q}-{\vec k}|}-2 \tilde V^{|{\vec q}|} \right)\tilde V^{|{\vec q}|} \notag\\
& \hspace*{1cm} \times  \left[    f^{0}_{\vec k} f^{0}_{\vec{k'}} f^{-}_{\vec{k'}+\vec{q}} f^{-}_{\vec{k}-\vec{q}} 
                      -  f^{0}_{\vec{k'}+\vec{q}} f^{0}_{\vec{k}-\vec{q}} f^{-}_{{\vec k'}} f^{-}_{{\vec k}}
                \right],  \label{eq:dev}
\end{align}
%
with the abbreviation $f^{-}_{\vec{k}} = 1- f^{0}_{\vec{k}}$. 
%
Equation~\eqref{eq:dev} illustrates the interplay of electrons within one subband to renormalize the equilibrium Fermi function: Two electrons with wave vectors $\vec{k'}+\vec{q}, \vec{k}-\vec{q}$ are annihilated, while two electrons with $\vec{k}, \vec{k'}$ are created, respectively. Again, the phenomenological damping $\gamma$ represents higher order correlations.\cite{Schilp:PhysRevB:94}

\begin{figure}[t]
\centering
\includegraphics[clip, width=0.8\columnwidth]{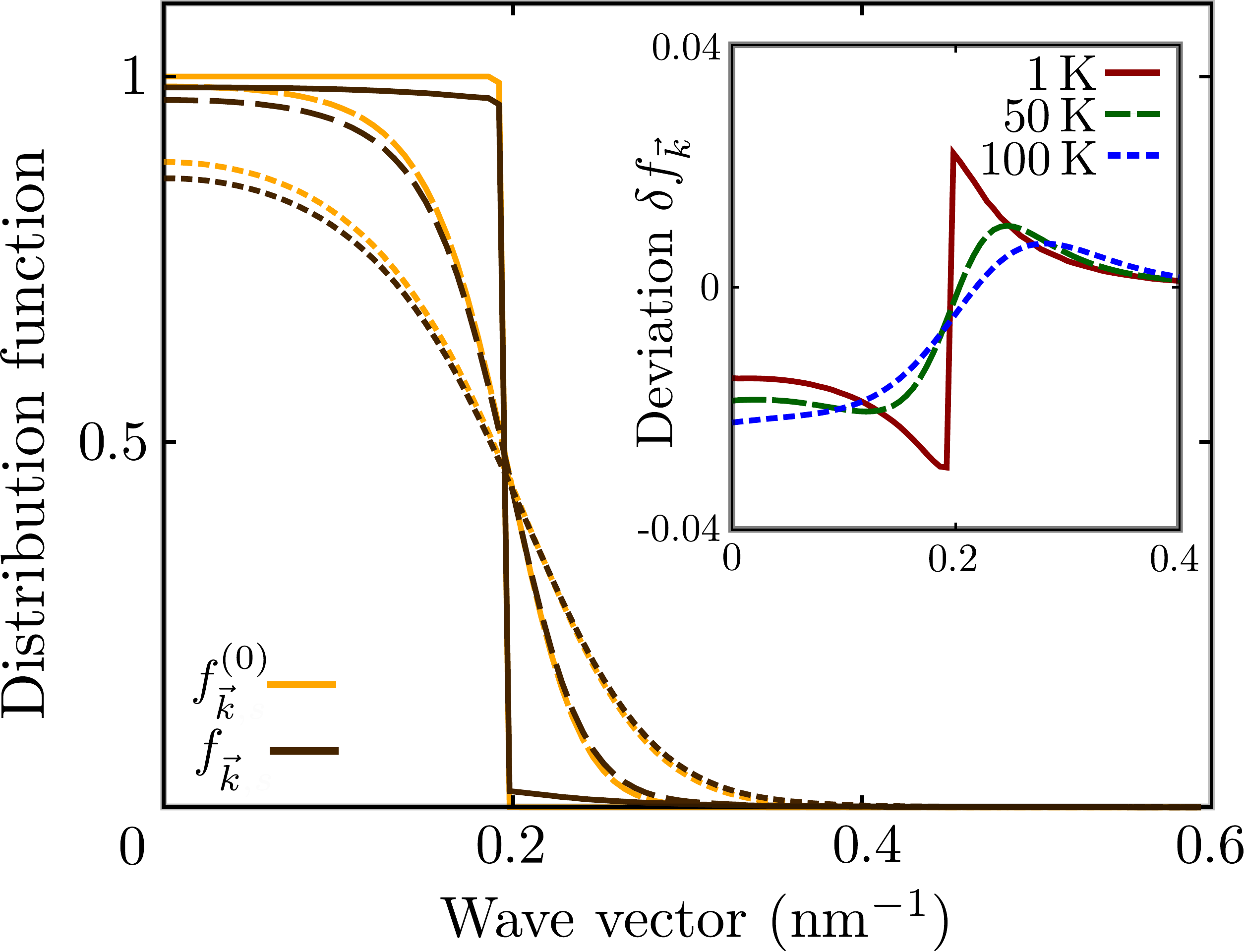}
\caption{(Color online) Electron distribution function of the electronic ground-state including (dark lines) and neglecting (light lines) Coulomb correlations for a 5 nm quantum well with an electronic doping density of $n_\text{dop} = 6.0 \times 10^{11}$ cm$^{-2}$ and various temperatures. Inset: Corresponding deviation $\delta f_{ {\vec k}} = f_{\vec k} - f^{(0)}_{\vec k}$ from the Fermi functions.}
\label{fig:corrFermi}
\end{figure}

Our result for the correlated ground-state distribution function $f_{\vec k} $ is plotted in Fig.~\ref{fig:corrFermi} for three different temperatures: $T$ = 1 K (solid lines), 50 K (dashed lines), and 100 K (dotted lines).
Comparing $f_{\vec k}$ (dark lines) with the equilibrium Fermi distribution $f^{(0)}_{\vec k}$ (light lines), we observe a slight decrease (increase) of the electron distribution for wave vectors below (above) the Fermi edge,
already known from electron gases in metals.\cite{Takada:PhysRevB:91,Daniel:PhysRev:60} This is especially pronounced for $T$ = 1 K, where one now finds available states below and a finite population above the Fermi edge.
Looking at the deviation from the Fermi function $\delta f_{ {\vec k}}$ (plotted in the inset of Fig.~\ref{fig:corrFermi}), one can see the sharp edge around the Fermi energy and a renormalization up to $3\%$ at a temperature of $T = 1$~K. The observed features soften for higher temperatures but remain on the same order of magnitude.

Even though the total deviation from the ideal Fermi function does not change strongly with temperature, we expect the influence of the correlations on scattering processes to decrease for rising temperatures: (i) The relative importance of the allowed scattering processes decreases with an increasing softening of the Fermi edge (see Fig.~\ref{fig:corrFermi}) and, at the same time, (ii) the deviation from the ideal Fermi function $\delta f_{ {\vec k}}$ decreases close to the Fermi edge. The scattering processes which go with the Coulomb coupling $\tilde{V}^{q} \sim 1/ |\vec q|$ are responsible for the heigth of $\delta f_{ {\vec k}}$. Since for larger temperatures the electron scattering is less common around the Fermi edge, $\delta f_{ {\vec k}}$ also decreases.

Even if the calculated $\delta f_{\vec k}$ is quantitatively quite small, it opens up a new physical scenario for the low-temperature regime: $\delta f_{\vec k} \neq 0$ leads to an occupation above and to available states below the Fermi edge, in particular for temperatures near 0~K, and  allows scattering which is otherwise prohibited due to Pauli blocking. 

\begin{figure}[t]
\centering
\includegraphics[clip, width=\columnwidth]{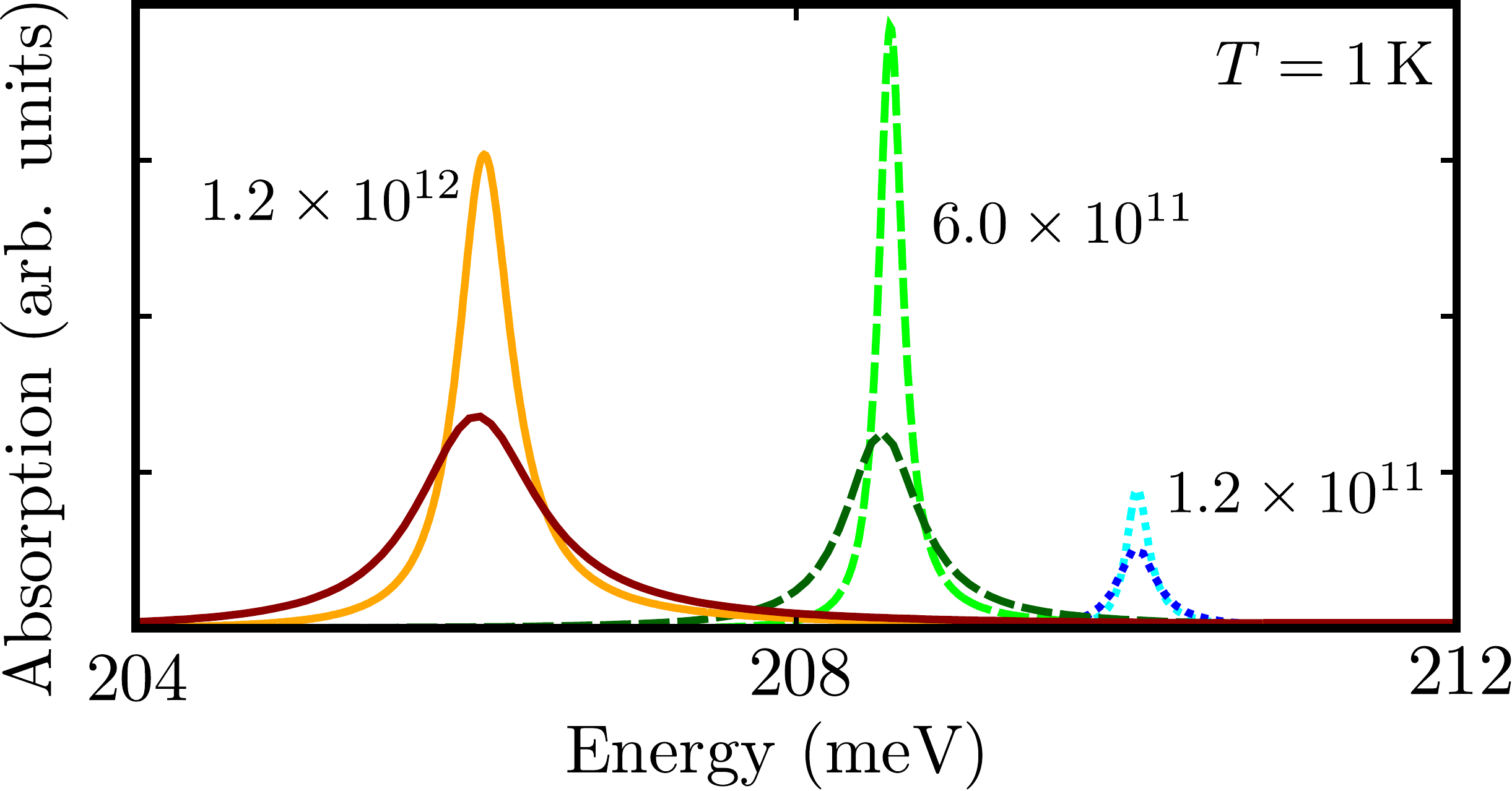}
\caption{(Color online) ISB absorption spectrum of a 5\,nm quantum well at a temperature $T$ = 1~K and different doping densities $n_\text{dop}$ (in $\text{cm}^{-2}$) including (dark lines) and neglecting (light lines) the electronic ground-state correlations.} \label{fig:var_n}
\end{figure}

Besides the dependence on the temperature, the observed correlation effects also depend on the quantum well width, where a slightly stronger deviation is found for smaller well widths (not shown). This is due to the stronger Coulomb coupling between the electrons for decreasing well widths. The doping density also influences the magnitude of $\delta f_{\vec k}$, where an optimal value of the density leads to a maximal deviation (cf. also the absorption spectra in Fig.~\ref{fig:var_n}). The decrease of the deviation for smaller densities can be explained by the short-range nature of the electronic correlations: for a larger mean-free path (Wigner-Seitz radius) $r_s$ between the electrons,\cite{Ziman::92} the probability of a momentum transfer decreases. On the other hand, we expect a decreasing deviation at a certain point for large densities since the the kinetic energy, which goes with $1/{r_s}^2$, overweigths the Coulomb interaction, going with $1/r_s$ between the electrons.\cite{Mahan::00}

(III) {\it Absorption spectra.}  
Next, to discuss experimental observables, the influence of the ground-state correlations on the linear quantum well intersubband absorption spectrum is studied. 
Figure~\ref{fig:var_n} shows the spectrum calculated from the dynamics of the polarization given by Eq.~(\ref{eq:polariz}) for different doping densities at a temperature of $T$ = 1~K. Here, the influence of the electronic ground-state correlations is clearly visible.
We find a strong broadening of the absorption line shape including ground-state correlations (dark lines) compared to the spectrum neglecting the electronic correlations (light lines). The full width at half maximum in the former is up to three times larger than in the latter case. In addition, the calculation including the correlated ground state shows a small energetic renormalization.
\begin{figure}
\centering
\includegraphics[clip, width=\columnwidth]{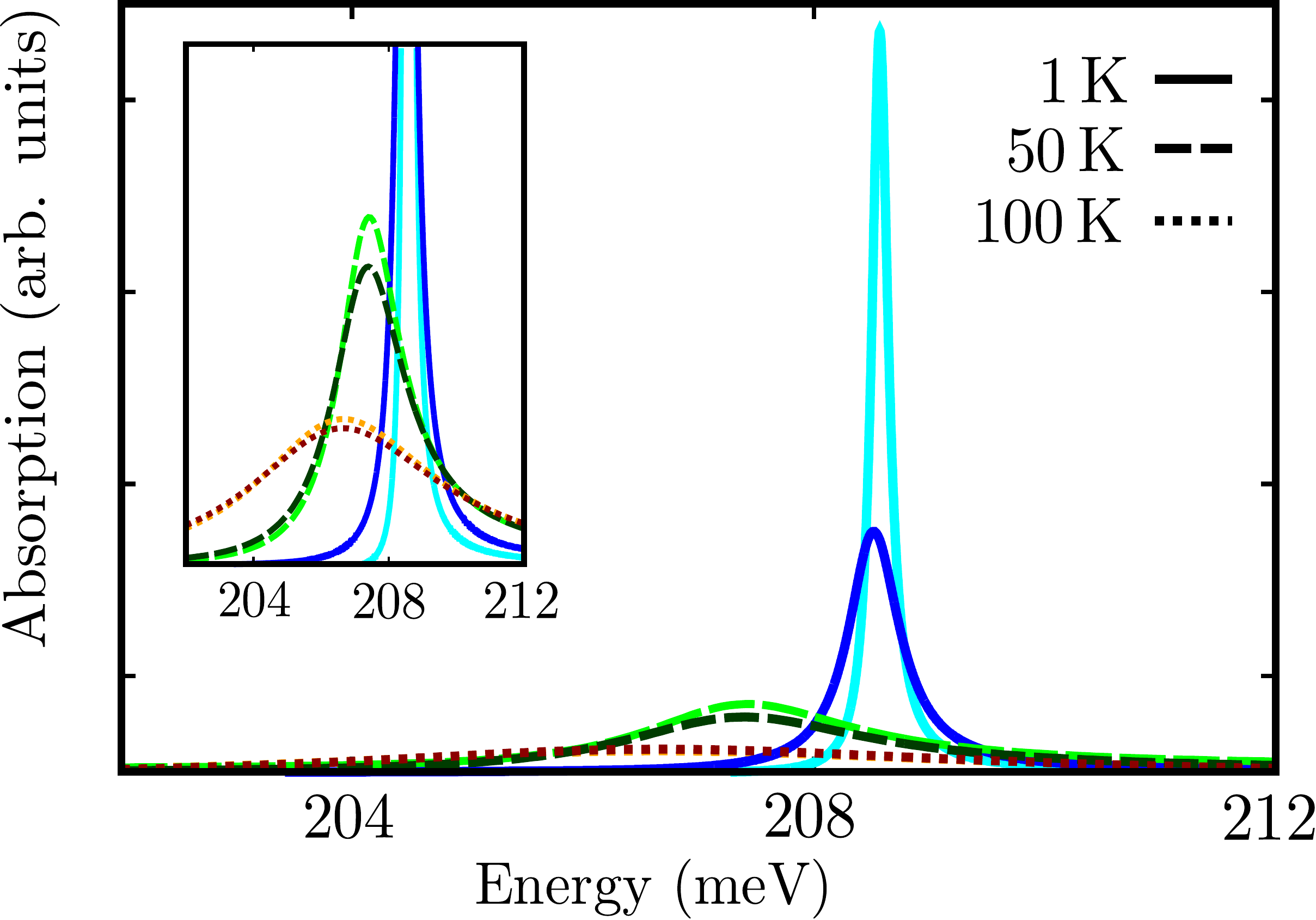}
\caption{(Color online) ISB absorption spectrum of a 5~nm quantum well with $n_\text{dop} = 6.0 \times 10^{11}~\text{cm}^{-2}$ including (dark lines) and neglecting (light lines) the electronic ground-state correlations for different temperatures. Inset: Enlarged view of the absorption line shapes.}
\label{fig:onlyscatt}
\end{figure}
This substantiates our prior claim that even small ground-state correlations have a strong influence at low temperatures: The absorption linewidth increases due to the generation of available scattering states, reflecting a partial cancellation of the strong Pauli blocking by the electronic correlations.
For the parameters used here, the difference between the absorption spectrum including and neglecting ground-state correlations is maximal for a doping density of $n_\text{dop} = 6.0\times10^{11} \text{cm}^{-2}$. The shift to lower energies for increasing densities is mainly due to the energetic renormalization resulting from the Hartree-Fock contributions,\cite{Li:PhysRevB:04,Waldmuller:PhysRevB:04} leading to a smaller effective band gap between the lower and upper subband.
Since we want to focus on the impact of the ground-state correlations on the absorption spectrum, we will restrict the following investigations to a 5 nm quantum well using a doping density of $n_\text{dop} = 6.0 \times 10^{11}\text{cm}^{-2}$.

In Fig.~\ref{fig:onlyscatt}, we show the absorption spectra for different temperatures. Comparing the results including (dark lines) and neglecting (light lines) ground-state correlations, we find that for temperatures higher than 50~K, the correlation effects are of no significant relevance as discussed earlier in the temperature dependence of $\delta f_{ {\vec k}}$ (cf. the inset of Fig.~\ref{fig:onlyscatt}). Instead, nonparabolicity effects gain importance, leading to an asymmetric line shape especially for smaller well widths due to the occupation of higher energetic states.\cite{Waldmuller:PhysRevB:04} For 50~K and below, the absorption spectrum shows a significant broadening leading to a strongly reduced peak absorption. Furthermore, the calculation with the correlated ground-state shows a small energetic renormalization. 

Although we find that the ground-state correlations lead to a significant broadening of the spectrum at low temperatures, the linewidth is still strongly underestimated compared to experimental findings.\cite{Kaindl::00,Shih:PhysRevB:05} In order to present a full and consistent description of ISB absorption including all important scattering processes which contribute to the absorption line shape, electron-phonon interaction must also be taken into account.\cite{Li:PhysRevB:04,Waldmuller:PhysRevB:04} This is assumed to be the major broadening mechanism for temperatures higher than $T$ =100~K.\cite{Butscher:PhysStatusSolidiB:04} For low temperatures, spontaneous phonon emission yields a temperature-independent contribution to the linewidth. 
\begin{figure}[h]
\centering
\includegraphics[clip,width=\columnwidth]{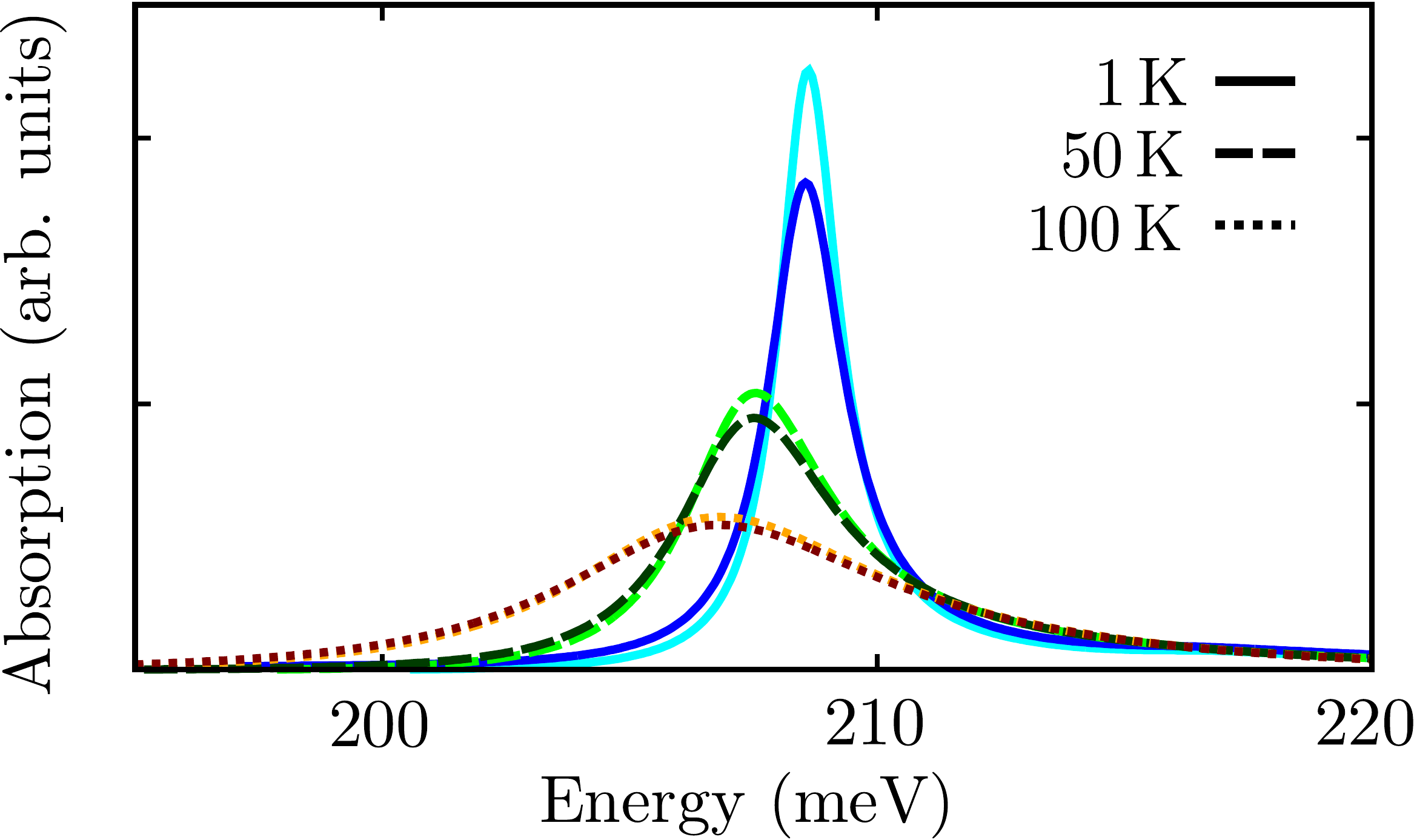}
\caption{(Color online) ISB absorption spectrum of a 5~nm quantum well including (dark lines) and neglecting (light lines) the electronic ground-state correlations, with microscopically calculated Markovian electron-phonon scattering rates for a doping density $n_\text{dop} = 6.0 \times 10^{11} \text{cm}^{-2}$ for different temperatures. 
}\label{fig:withphonon}
\end{figure}
We therefore extract temperature-dependent Markovian electron-phonon scattering rates from earlier microscopic calculations and include them as an additional 
dephasing contribution $\gamma_\text{phon}$ in the calculations.\cite{Butscher:PhysStatusSolidiB:04}
The resulting calculated ISB absorption spectra are found in Fig.~\ref{fig:withphonon}. We find that the influence of the ground-state correlations is strongly masked by the phonon-induced dephasing. Still, a deviation between the absorption spectra neglecting ground-state correlations (light solid line) and including ground-state correlations (dark solid line) can be found for temperatures lower than 50~K, where the the spectra neglecting ground-state correlations show less broadening than the ones including ground-state correlations. For $T$ = 100~K, the difference is hardly visible.

In {\it conclusion}, we have presented a microscopic theory for the description of ISB quantum well absorption including electron-electron contributions and electronic ground-state correlations. We showed that by including ground-state correlations, the absorption linewidth for temperatures $T <$ 50~K shows significant broadening, where the full width at half maximum is increased by up to a factor 3 at $T$ = 1~K. The additional inclusion of electron-phonon scattering masks the impact of the ground-state correlations on the spectral width. 
In a next step, a more consistent treatment of both the electron-phonon and electron-electron dephasing, including full non-Markovian effects,\cite{Butscher:PhysRevLett:06} should be carried out to allow deeper insights into the influence of electronic ground-state correlations.

\begin{acknowledgments}
We acknowledge financial support by the Deutsche Forschungsgemeinschaft (DFG) through the project KN 427/4-1 and the Alexander von Humboldt Foundation through the Feodor-Lynen program. 
\end{acknowledgments}

\end{document}